\documentstyle[emulateapj]{article}

\def\lsim{\mathrel{\rlap{\lower 4pt \hbox{\hskip 1pt $\sim$}}\raise 1pt \hbox
        {$<$}}}
\def\gsim{\mathrel{\rlap{\lower 4pt \hbox{\hskip 1pt $\sim$}}\raise 1pt \hbox
        {$>$}}}

\lefthead{Nakamura et al.}
\righthead{Nucleosynthesis in Type II supernovae and
the abundances in metal-poor stars}

\begin{document}

\submitted{To appear in the Astrophysical Journal}

\title{Nucleosynthesis in Type II supernovae and\\
the abundances in metal-poor stars}

\author{Takayoshi Nakamura,
Hideyuki Umeda\altaffilmark{1}, and Ken'ichi Nomoto\altaffilmark{1,2}}
\affil{Department of Astronomy, School of Science, University of Tokyo, Japan}

\and

\author{Friedrich-Karl Thielemann\altaffilmark{2}}
\affil{Department f\"ur Physik und Astronomie,
Universit\"at Basel, Switzerland}

\and

\author{Adam Burrows\altaffilmark{2}}
\affil{Department of Astronomy,
The University of Arizona, Tucson, AZ 85721, USA}

\altaffiltext{1}{Research Center for the Early Universe,
University of Tokyo, Japan}
\altaffiltext{2}{Institute for Theoretical Physics,
University of California, Santa Barbara, USA}

\begin{abstract}
We explore the effects on
nucleosynthesis in Type II supernovae
of various parameters (mass cut, neutron excess,
explosion energy, progenitor mass) in order to
explain the observed trends of the iron-peak element
abundance ratios ([Cr/Fe], [Mn/Fe], [Co/Fe] and [Ni/Fe])
in halo stars as a function of metallicity
for the range $ -4 \le$ [Fe/H] $\le -2.5$.
[Cr/Fe] and [Mn/Fe] decrease with decreasing [Fe/H],
while [Co/Fe] behaves the opposite way and increases.
We show that such a behavior can be explained by a variation
of mass cuts in Type II supernovae as a function of progenitor mass, which
provides a changing mix of nucleosynthesis from an alpha-rich
freeze-out of Si-burning and incomplete Si-burning. This explanation is
consistent with the amount of ejected $^{56}$Ni determined from modeling
the early light curves of individual supernovae.
We also suggest that
the ratio [H/Fe] of halo stars
is mainly determined by the mass of interstellar hydrogen
mixed with the ejecta of a single supernova
which is larger for larger
explosion energy and
the larger Str\"omgren
radius of the progenitor.
\end{abstract}


\section{INTRODUCTION}\label{sec:intro}
\begin{figure}
\epsscale{.6}
\plotone{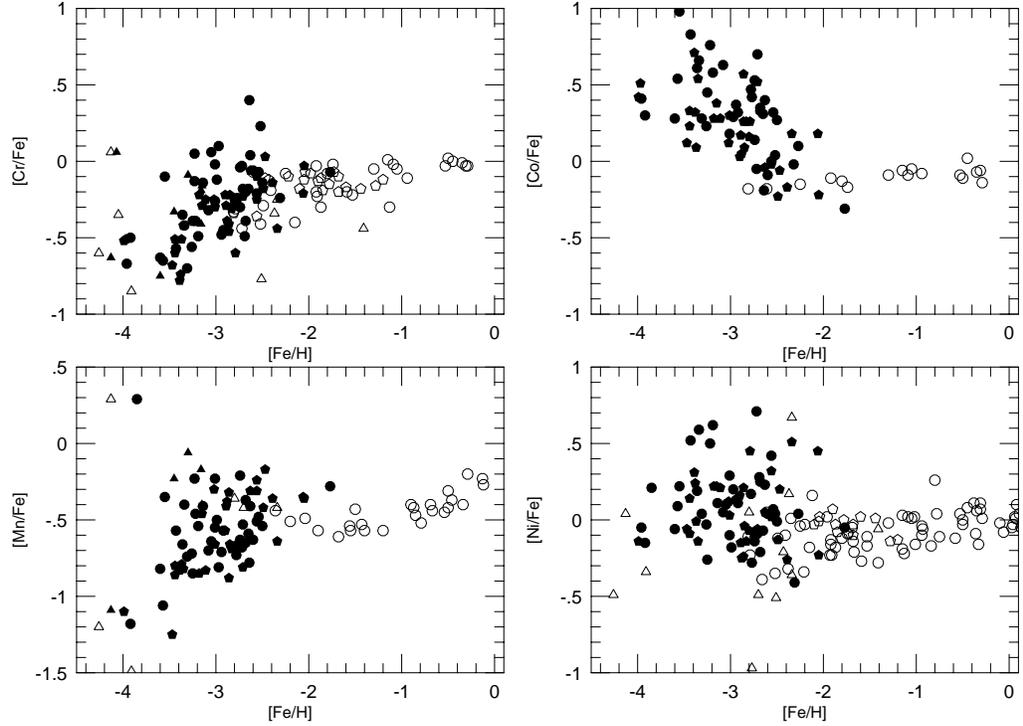}
\caption{
The trends of iron-peak abundance ratios.
These data are from Ryan et al. (1991, 1996) (filled circles);
McWilliam et al. (1995a), McWilliam (1998) (filled pentagons);
Norris et al. (1993), Primas et al. (1994) (filled triangles); 
Gratton \& Sneden (1987, 1988, 1991), and Gratton (1989) (open circles);
Magain (1989), and Zhao \& Magain (1990) (open pentagons); and
Molaro \& Bonifacio (1990), Molaro \& Castelli (1990),
Peterson et al. (1990) (open triangles).
\label{fig:newtrends}}
\end{figure}

Very metal poor stars provide important clues
to investigate early Galactic chemical and dynamical evolution
because they contain valuable information about this time.
We can learn how the Galaxy evolved chemically (as well as dynamically)
in its early phase from the observation of these stars.
Therefore, a large number of observations and abundance analyses
have been performed.
Until recently,
no significant abundance
change with metallicity had been observed.
However, recent high-resolution abundance surveys have discovered
interesting trends of [Cr/Fe], [Mn/Fe] and [Co/Fe]
with respect to [Fe/H]
([A/B] = log$_{10}$(A/B)$-$log$_{10}$(A/B)$_\odot$).
Both [Cr/Fe] and [Mn/Fe] decrease
with decreasing metallicity from [Fe/H] = $-2.4$ to $-4.0$,
while [Co/Fe] increases (Figure \ref{fig:newtrends})
(McWilliam et al. 1995a, 1995b; McWilliam 1997;
see also Ryan et al. 1996 and reference therein).

\placefigure{fig:newtrends}

Over the last few years, these trends have been the subject of controversy.
McWilliam et al. (1995b) discussed the time-delay mechanism,
which produces the abundance trends
by the time-delay between SNe II with different masses
or between SNe I and SNe II.
They argued that this mechanism
works for several species such as $\alpha$-elements
but has difficulties with the above trends.
McWilliam et al. (1995b) also pointed out that
the Galactic halo formation model proposed
by Searle \& Zinn (1978) is at odds with the time-delay mechanism.
Instead, they suggested that the metallicity-dependent
supernova progenitor mass function could explain these trends,
while still being consistent with Searle \& Zinn (1978).
Audouze \& Silk (1995) interpreted the observations
as evidence that the Galaxy was poorly mixed
and proposed that at most two or three supernovae could have
contaminated any particular cloud.
In favor of this hypothesis,
Ryan et al. (1996) proposed
that a variation of the supernova explosion energy,
which is connected with both the yields and
the amount of ejecta dilution,
could cause these trends.
McWilliam (1997) and Searle \& McWilliam (1998)
suggested that diluting a primordial composition
with solar composition supernova ejecta
could explain the trends.
In principle,
hypothetical pre-galactic population III objects,
such as very massive stars or
pair creation supernovae (e.g., Bond et al. 1984),
could be the cause of the observed behavior.
However, none of these objects has been investigated in full
detail in quantitative nucleosynthesis yield studies
nor ever been incorporated
in chemical evolution calculations.

Timmes et al. (1995) constructed a Galactic evolution model
using a grid of Type II supernova models
(Woosley \& Weaver 1995, hereafter WW95),
but did not explain the above-mentioned abundance trends.
In Timmes et al. (1995),
only the region with [Fe/H] $ > -3.0$ was investigated,
so that only limited information is available.
However, their results seem to suggest that
the metallicity effect alone cannot explain the observed trends.

The purpose of this paper is to provide a qualitative explanation
of the halo abundance data
within the framework of Type II supernovae (SNe II)
explosive nucleosynthesis. 
It should be pointed out that
self-consistent explosion and nucleosynthesis calculations, which start from
core collapse, have not yet successfully determined the mass cut between the
central neutron star (or black hole) and the ejected envelope. Therefore,
the mass cut of SNe II as a function of the progenitor mass is still an open
question and can be treated as a free parameter. This investigation
focuses on the mass cut, which determines the amount of $^{56}$Ni
in SN II ejecta, and its possible connection with the amount of 
(stable) Cr, Mn, Co, and Ni ejected.
The connection between stellar evolution, explosive yields, and
galactic evolution, observed via abundances in low metallicity stars,
is made by performing chemical evolution studies in the framework previously
described by Tsujimoto et al. (1995, 1997).

Although we are investigating the nucleosynthesis of low-metallicity stars,
we use solar metallicity pre-supernova models in this paper
for the following reasons. Previously, only WW95 calculated
nucleosynthesis of SNe II using progenitor models with low metallicities;
however, their progenitor models have not yet been published. 
Their results suggest that although the yields of iron peak elements
depend on the progenitor metallicity, their variation in the 
range $Z = (10^{-2} - 10^{-4}) Z_\odot$ is only about 20\% or less.
Since we are interested only in $Z$ in this range, 
we neglect metallicity effects on progenitors and explore
the dependence on
other parameters, such as the stellar mass,
mass cut, explosion energy, and neutron excess ($= 1 - 2Y_e$,
where $Y_e$ denotes the electron mole number).  
With this procedure, we can investigate the trends of the yields, i.e.,
how much the yields increase or decrease with these parameters.

In \S \ref{sec:nsinSNII}, we review
the nucleosynthesis yields of SNe II (including
their dependence on the mass cut, explosion energy,
and $Y_e$ of the innermost ejected matter).
The results are applied in chemical evolution calculations
in \S \ref{sec:abundhalo}
in order to explore whether the observed abundance patterns
in the metal-poor stars
can be explained with our approach.
Section \ref{sec:discussions}
contains a more in-depth discussion,
listing both the successes and the remaining open questions.

\section{NUCLEOSYNTHESIS IN TYPE II SUPERNOVAE}\label{sec:nsinSNII}

\subsection{Models}

Stars initially more massive than 8$M_\odot$ explode
as SNe II at the end of their life,
leaving neutron stars or black holes.
When they explode, nucleosynthesis proceeds explosively
because of high temperatures and
densities in the deep stellar interior.
In order to investigate the explosive nucleosynthesis of SNe II,
we solve nuclear reaction networks,
together with hydrodynamical equations,
and calculate the energy generation by explosive burning.

Our calculations are performed in two steps.
The first step is a hydrodynamical simulation of the SN II
explosions with a small nuclear reaction network
which contains only 13 alpha nuclei
($^{4}$He, $^{12}$C, $^{16}$O, $^{20}$Ne, $^{24}$Mg, $^{28}$Si,
$^{32}$S, $^{36}$Ar, $^{40}$Ca, $^{44}$Ti,
$^{48}$Cr, $^{52}$Fe, and $^{56}$Ni).
The hydrodynamical simulations are carried out 
with a one dimensional PPM (piecewise parabolic method) code
(Colella and Woodward 1984).
We generate a shock by depositing thermal energy
below the mass cut that divides the ejecta and the collapsing core,
and perform the induced nucleosynthesis 
calculations following the procedure of our previous  investigations
(e.g. Hashimoto et al. 1989;
Thielemann, Hashimoto, \& Nomoto 1990; Thielemann,
Nomoto, \& Hashimoto 1996, hereafter TNH96).
In the second step,
at each mesh point of the hydrodynamical model,
post-processing calculations are performed
with an extended reaction network (Hix \& Thielemann 1996),
which contains 211 isotopes,
and provides precise total yields (also for minor abundances).

\begin{figure}
\epsscale{.6}
\plotone{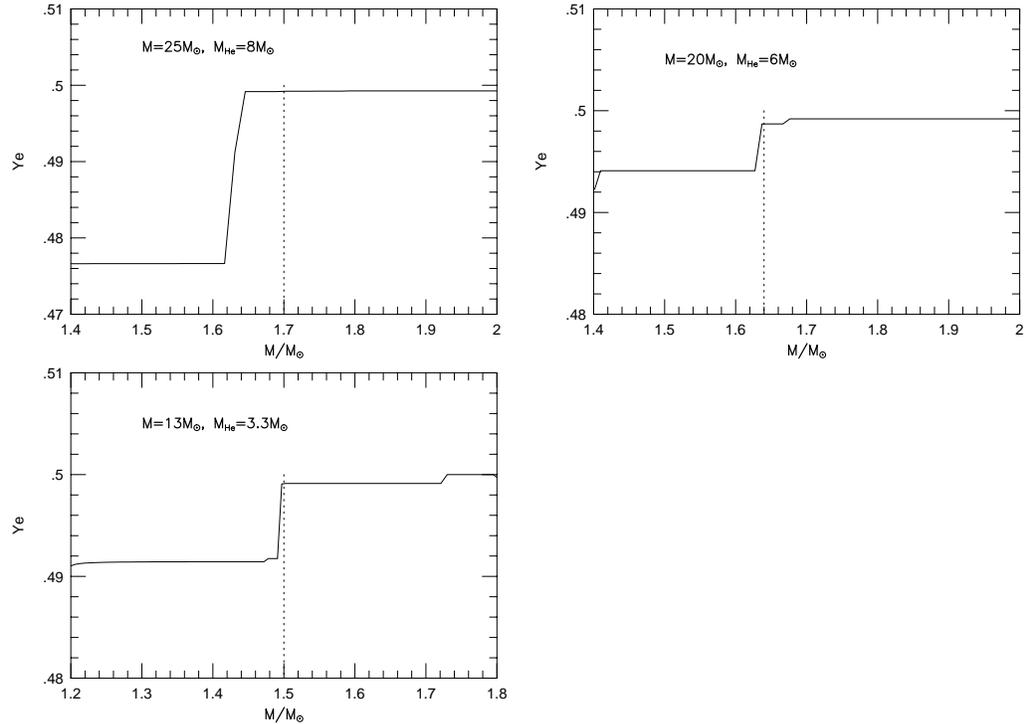}
\caption{
$Y_{e}$ profiles of the models in NH88.
Dotted lines indicate the boundaries of $Y_{e}$ modification.
In our calculations, $Y_{e}$ is modified to be a constant value
below these lines,
while the NH88's $Y_{e}$ distribution of the progenitor models
is maintained above these lines.
\label{fig:YeNH88}}
\end{figure}

In our calculation, the progenitor models are taken from
Nomoto \& Hashimoto (1988, hereafter NH88).
Figure \ref{fig:YeNH88} shows the $Y_{e}$ profiles of the models
(25$M_\odot$, 20$M_\odot$ and 13$M_\odot$) in NH88.
In this paper, $Y_{e}$ in the deep stellar interior
is modified to take a constant value of $Y_{e}^{\rm deep}$
from $M_{\rm r} = M_{\rm cut}$ to 
$M_{\rm r} = 1.70M_\odot$ (for the 25$M_\odot$ star),
$M_{\rm r} = 1.64M_\odot$ (20$M_\odot$ star) and
$M_{\rm r} = 1.49M_\odot$ (13$M_\odot$ star).
NH88's $Y_{e}$ distribution of the progenitor
is adopted for the outer region above the 
boundaries shown in Figure \ref{fig:YeNH88}.
Some of the adopted mass cuts
are smaller than the Fe core masses obtained in
the stellar evolution calculations (NH88).
Note, however, that the Fe core mass,
the $Y_{e}$ profiles, and $Y_{e}^{\rm deep}$ are
subject to uncertainties involved in the stellar evolution calculations,
such as the treatment of convection, reactions rates
(especially $^{12}$C($\alpha,\gamma$)$^{16}$O)), etc.
Using the modified models,
we analyze the dependences on the 
mass cut, $Y_{e}$, etc, and obtain some constraints on these parameters.
We also treat mass cuts and explosion energies independently,
though these are physically related,
because of uncertainties in the explosion mechanism.
When analyzing the yields,
we do not include grid zones inside the mass cut
in the ejecta by assuming that
materials in these regions fall back onto a neutron star or a black hole.
The change in the final total energy caused by this treatment
is negligible.

\placefigure{fig:YeNH88}

In the following subsections, we show the dependence of SNe II yields
on various parameters (mass cut, neutron excess,
explosion energy, progenitor mass).
Among them,
the mass cut seems to be the most important parameter.

\subsection{Production of iron-group elements in Type II supernovae}

\begin{figure}
\epsscale{.6}
\plotone{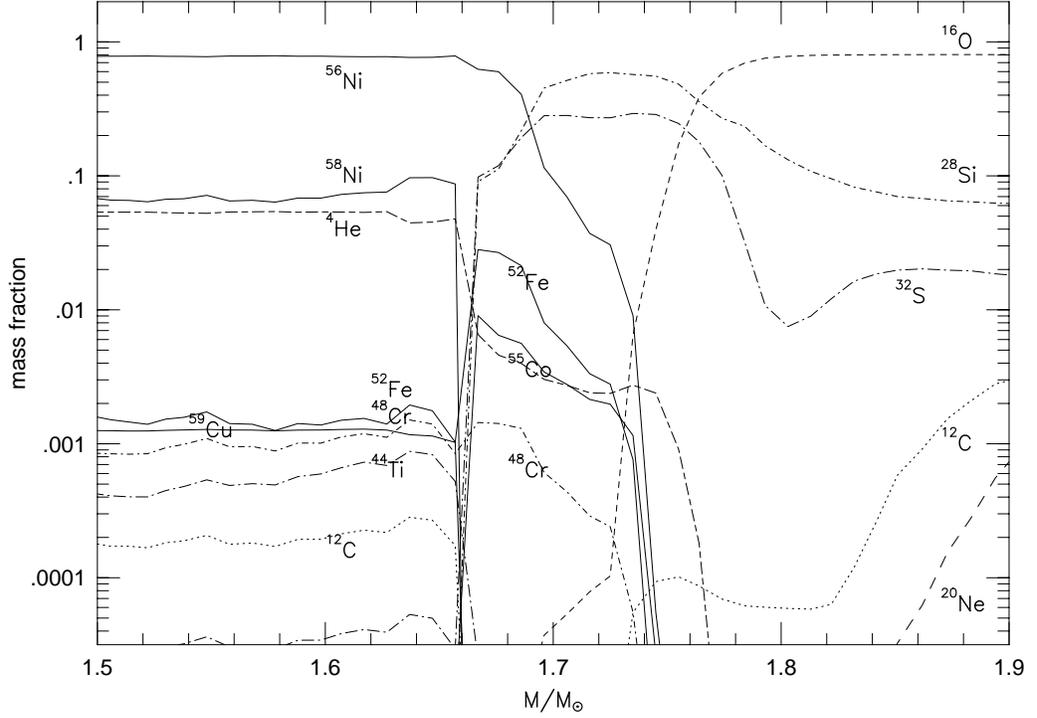}
\caption{
The isotopic composition of the ejecta from 20$M_\odot$ stars
(6$M_\odot$ He core).
\label{fig:sn6m10c13y8.elemp}}
\end{figure}

\placefigure{fig:sn6m10c13y8.elemp}

Figure \ref{fig:sn6m10c13y8.elemp} shows the isotopic composition
of the ejecta of
the explosion of the 20$M_\odot$ star as a typical SN II.
Here three distinctive regions are seen.
One is the innermost high temperature region.
The outer boundary of this region is determined by the condition that maximum
temperatures of $T \gsim 5\times 10^9$K is attained behind the shock.
In this region, nuclear
statistical equilibrium (NSE) is achieved
except for the slow triple-alpha process ($\alpha$-rich freezeout).
Thus, the nucleosynthesis depends only on $Y_e$ and the entropy,
being dominated by iron group elements
(Woosley, Arnett, \& Clayton 1973; Thielemann, Hashimoto, \& Nomoto 1990).
The most abundant element
in this region of complete Si-burning with alpha-rich freeze-out is
$^{56}$Ni, provided $Y_e > 0.49$. The second region,
which does not experience such high temperatures, undergoes incomplete
Si-burning with only partial Si exhaustion;
here the most abundant nucleus
changes from $^{56}$Ni to $^{28}$Si.
In the third region the temperatures are too low to produce any iron
group elements, and only explosive O, Ne, or C-burning take place.
Farther out in radius, regions are encountered where explosive nucleosynthesis
barely affects the pre-explosive composition and matter as produced
during the quasi-static evolution is ejected essentially unchanged.

In order to investigate the ratios [Cr/Fe], [Mn/Fe] and [Co/Fe]
we look into the regions where these elements are produced.
First of all, $^{56}$Ni, which decays into 
the most abundant Fe isotope $^{56}$Fe,
is produced not only in the complete Si-burning region
but also in the incomplete Si-burning region.
$^{52}$Fe and $^{55}$Co, which decay into $^{52}$Cr and $^{55}$Mn,
respectively, are mostly synthesized
in the incomplete Si-burning region.
$^{52}$Fe is also synthesized in the complete Si-burning region,
but not a large amount.
(Note that the ordinate of Figure \ref{fig:sn6m10c13y8.elemp}
is log-scaled.)
On the contrary, $^{59}$Cu, which decays into $^{59}$Co, is produced in
the complete Si-burning region. Note that $^{55}$Mn and
$^{59}$Co are the only stable isotopes of these elements and therefore,
their abundances are identical with the Mn and Co element abundances.
$^{52}$Cr dominates the
Cr element abundance by 84\% (in solar composition). Thus, it is
sufficient to take these isotopes into consideration when discussing the
abundances of [Cr, Mn, Co/Fe].

\subsection{Dependence on mass cuts}\label{subsec:donmc}

The above discussion
suggests that the choice of the mass cut can affect the ratios
[Cr/Fe], [Mn/Fe], and [Co/Fe]. 
For a deeper mass cut (i.e. smaller $M_{\rm cut}$),
the ejected mass of the complete Si-burning region is larger
(i.e., the masses of Fe and Co are larger),
while the ejected mass of the incomplete Si-burning region
remains the same
(i.e., the masses of Cr, Mn, and Fe are the same),
accordingly the ratios of [Cr/Fe] and 
[Mn/Fe] are smaller and [Co/Fe] is larger.
For a mass cut at larger radii
(larger $M_{\rm cut}$) these ratios show the opposite tendency.
Therefore, specific choices of mass cuts in SNe II might
explain the behavior of [Cr/Fe], [Mn/Fe], and [Co/Fe] in the metallicity
range $-4\le$ [Fe/H] $\le-2.5$.

\begin{figure}
\epsscale{.6}
\plotone{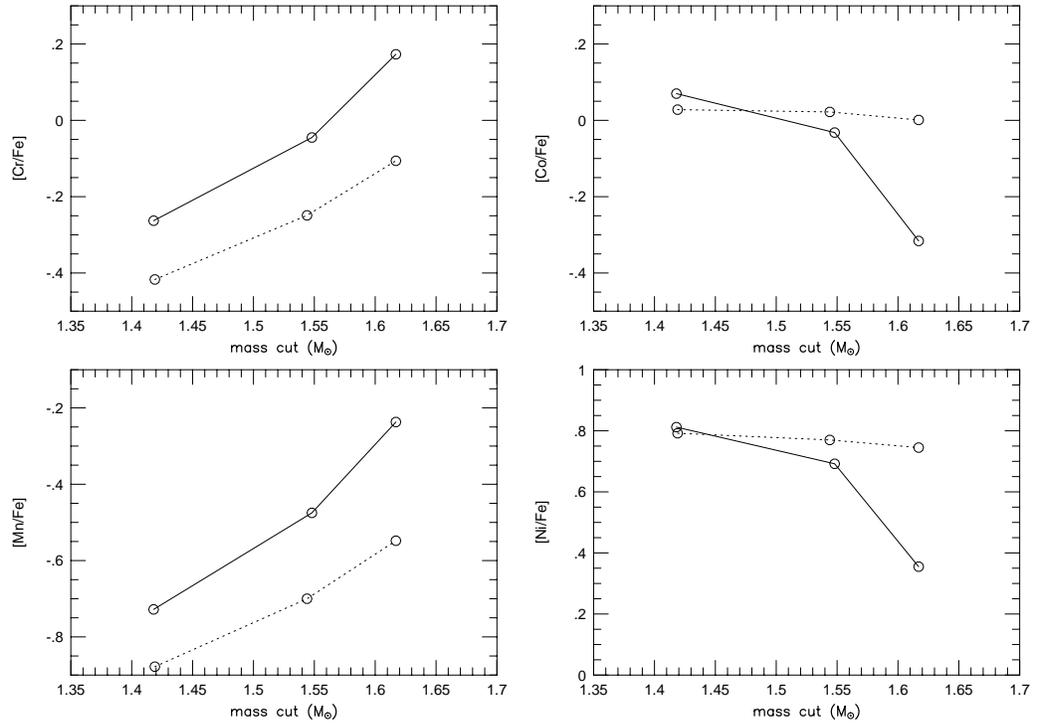}
\caption{
Dependence on mass cut.
Three models with different mass cuts are used for each line.
Solid line indicates supernovae of $M_{{\rm core}} = 6M_{\odot}$,
${\rm E}_{\rm exp} = 1.0 \times 10^{51}$ergs, $Y_{e}^{\rm deep}$ = 0.4940.
Dotted line indicates that of $M_{{\rm core}} = 8M_{\odot}$,
${\rm E}_{\rm exp} = 1.0 \times 10^{51}$ergs, $Y_{e}^{\rm deep}$ = 0.4985.
\label{fig:donmc4}}
\end{figure}

Table \ref{masscutdt} summarizes the yields
of the 25, 20 $M_\odot$ stars for various mass cuts.
Figure \ref{fig:donmc4} shows the dependence
of the abundance ratios on the mass cut.
Here we use the solar abundances by Anders and Grevesse (1989).
For smaller $M_{\rm cut}$, [Cr/Fe] and [Mn/Fe] are smaller,
while [Co/Fe] and [Ni/Fe] are larger.
Note that stable Ni is dominated by $^{58}$Ni and, thus,
[Ni/Fe] is dominated
by the $^{58}$Ni/$^{56}$Ni ratio.
One finds substantial
$^{58}$Ni production only in the inner complete Si-burning region,
so that the behavior of [Ni/Fe] is similar to [Co/Fe].
The observational trends
in Cr, Mn, Co and Ni are reproduced simultaneously.

\placetable{masscutdt}
\placefigure{fig:donmc4}

\subsection{Dependence on neutron excess}\label{subsec:Yedep}

\begin{figure}
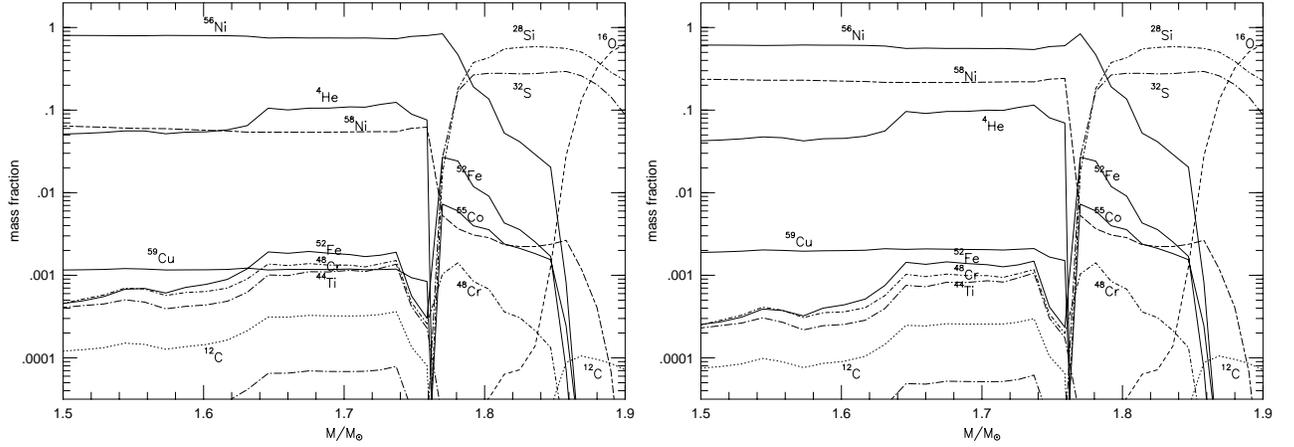

\epsscale{.37}
\plotone{fig5afig8b.epsi}
\hspace{1.5cm}
\plotone{fig5b.epsi}
\caption{
Dependence on $Y_{e}$.
The progenitor is the 8$M_\odot$ He core.
The explosion energy is $1.0 \times 10^{51}$ ergs.
For both models, $Y_{e}^{\rm deep}$ is modified to be constant,
0.4985 (left) and 0.4950 (right).
\label{fig:yedependence}}
\end{figure}

\placefigure{fig:yedependence}

\begin{figure}
\epsscale{.6}
\plotone{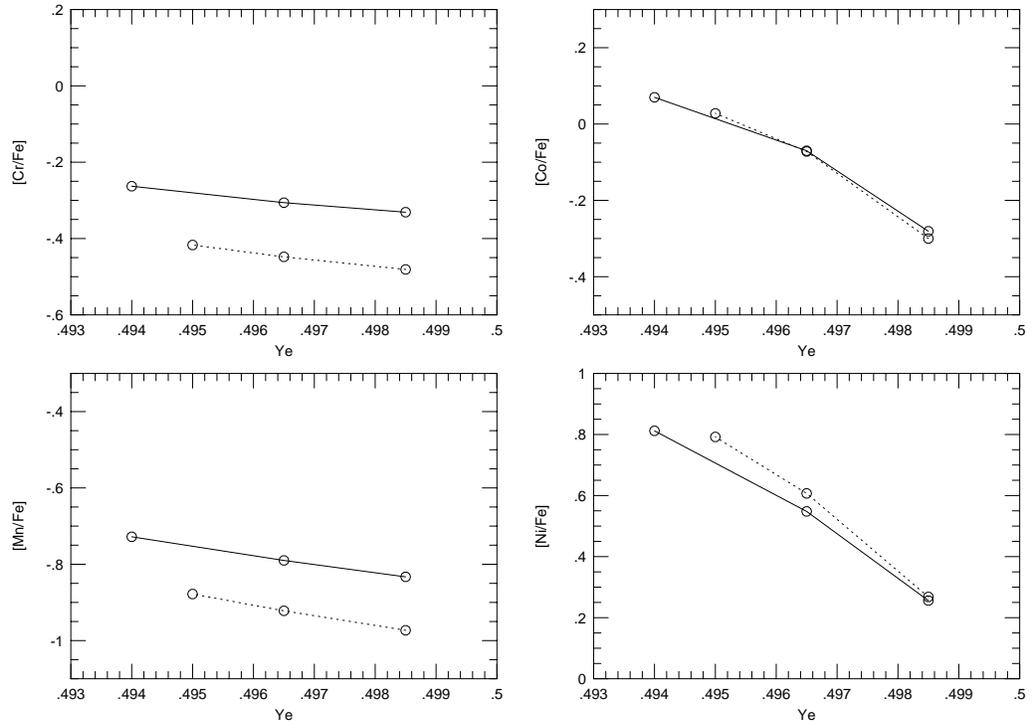}
\caption{
Dependence on $Y_{e}$.
Three models with different $Y_{e}$ are used for each line.
Solid line indicates supernovae of $M_{{\rm core}} = 6M_{\odot}$,
${\rm E}_{\rm exp} = 1.0 \times 10^{51}$ergs,
$M_{{\rm cut}} = 1.418M_{\odot}$.
Dotted line indicates $M_{{\rm core}} = 8M_{\odot}$,
${\rm E}_{\rm exp} = 1.0 \times 10^{51}$ergs,
$M_{{\rm cut}} = 1.419M_{\odot}$.
\label{fig:donYe4}}
\end{figure}

Figure \ref{fig:yedependence} shows the $Y_{e}$ dependence
of SN II yields for $Y_{e}^{\rm deep}$ = 0.4985 (left) and 0.4950 (right).
The progenitor is
the 25$M_\odot$ star with
the 8$M_\odot$ He core and
the explosion energy is $1.0 \times 10^{51}$ ergs.
Table \ref{yedt} summarizes the nucleosynthesis products and
compares the abundance ratios with the solar ratios
for different values of $Y_{e}^{\rm deep}$.
Figure \ref{fig:donYe4} shows
how these ratios depend on $Y_{e}$.
For smaller $Y_{e}$
(i.e., more neutron-rich environment),
more $^{59}$Cu
(which decays into $^{59}$Co) and $^{58}$Ni are produced,
while the yield of $^{52}$Fe is smaller.
$^{58}$Ni is more sensitive to $Y_{e}$ than $^{59}$Co.
The $^{56}$Ni yield is smaller for smaller $Y_{e}$.
Thus, all the metal to iron ratios are larger for smaller $Y_{e}$.

\placetable{yedt}
\placefigure{fig:donYe4}

\subsection{Dependence on explosion energy}

\begin{figure}
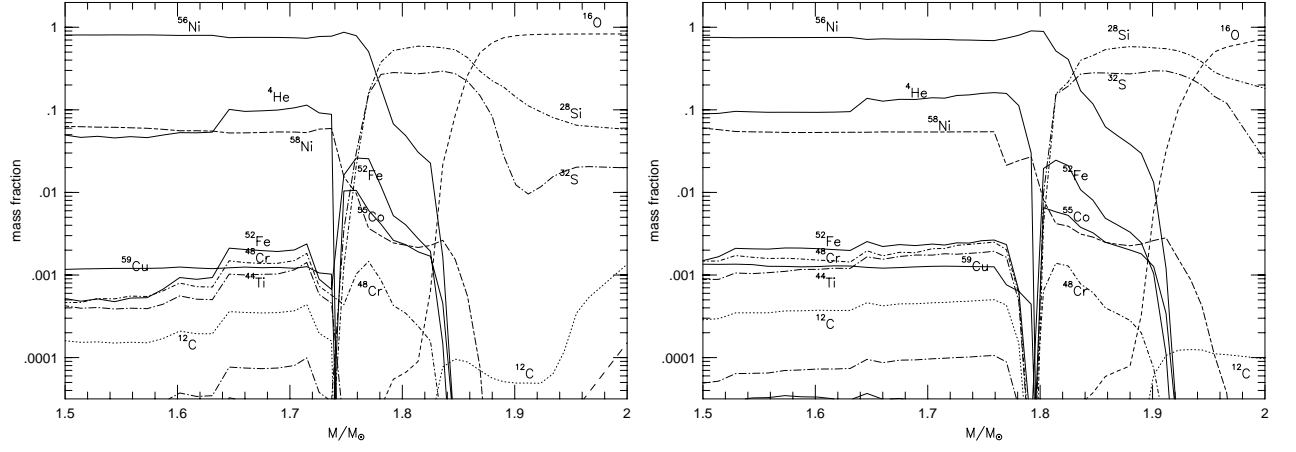

\epsscale{.37}
\plotone{fig7a.epsi}
\hspace{1.5cm}
\plotone{fig7b.epsi}
\caption{Dependence on explosion energy.
They have the same progenitor the 8M$_{\odot}$ He core
and the $Y_{e}^{\rm deep}$ of 0.4985,
but the explosion energies are $0.4 \times 10^{51}$ergs (left)
and $2.0 \times 10^{51}$ergs (right).
\label{fig:energydependence}}
\end{figure}

\placefigure{fig:energydependence}

The explosion energy of a typical SN II
is considered to be $E \sim 1 \times 10^{51}$ ergs.
For example, the explosion energy of SN 1987A has been estimated to be
$E = (1.0-1.5)\times 10^{51}$ ergs from modeling of the early light curve
(Shigeyama et al. 1987, 1988, 1990;
Woosley et al. 1988; Nomoto et al. 1997; Nakamura et al. 1998).
Figure \ref{fig:energydependence} shows the energy dependence
of Type II supernova yields for the same progenitor model, i.e.,
the $8M_\odot$ He core of the $25M_\odot$ star.
Here $Y_{e}^{\rm deep} = 0.4985$,
while E = $0.4 \times 10^{51}$ergs (left)
and $2.0 \times 10^{51}$ergs (right).

It is seen that the larger explosion energy forms
a larger region of complete silicon burning,
while the incomplete silicon burning region is enlarged only slightly.
In other words, for larger $E$,
the region of incomplete silicon burning is
shifted outward in mass
without an appreciable change of the enclosed mass.
Therefore, the SN II with a larger explosion energy produces
a larger amount of iron group elements
if mass cuts are the same.
That is, a larger explosion energy affects abundance ratios
in the same way as a smaller mass cut.
The explosion energy of SNe II can vary
by a factor of five (Burrows 1998).
We find that changing explosion energies
from $0.4 \times 10^{51}$ to $2.0 \times 10^{51}$ ergs
causes some changes in the abundance ratios,
but cannot explain the large variations in the observations.

\subsection{Dependence on progenitor mass}

\begin{figure}
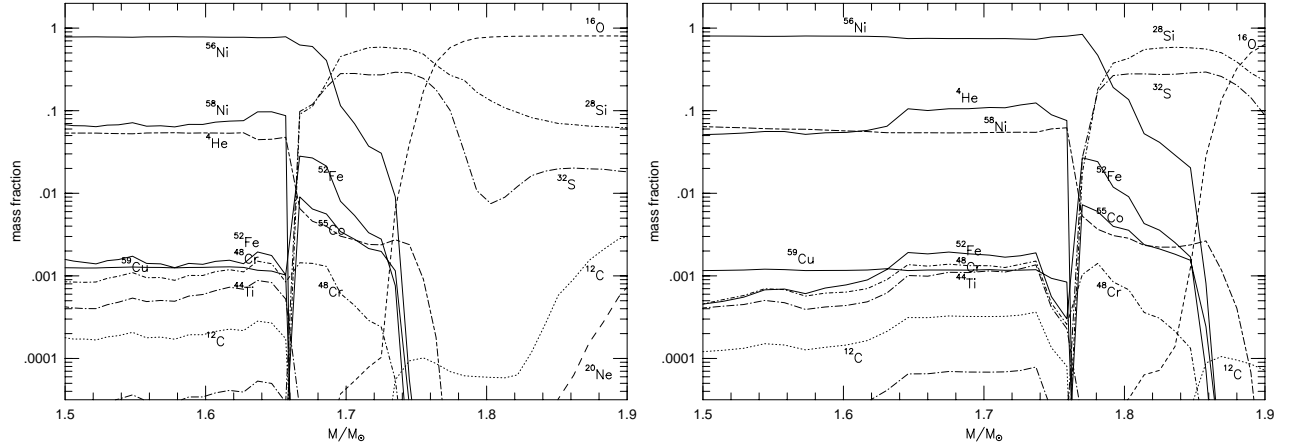

\epsscale{.37}
\plotone{fig3fig8a.epsi}
\hspace{1.5cm}
\plotone{fig5afig8b.epsi}
\caption{
Dependence on progenitor mass.
Both models have the same explosion energy of $1.0 \times 10^{51}$ ergs
and a $Y_{e}^{\rm deep}$ of 0.4985,
but different progenitors,
which are the 6$M_\odot$ He core of
the 20$M_\odot$ star (left) and
the 8$M_\odot$ He core of
the 25$M_\odot$ star (right).
\label{fig:pmdependence}}
\end{figure}

\placefigure{fig:pmdependence}

Figure \ref{fig:pmdependence} shows the progenitor mass dependence
of SN II nucleosynthesis.
The progenitors are the 6$M_\odot$ He core of
the 20$M_\odot$ star (left) and
the 8$M_\odot$ He core of
the 25$M_\odot$ star (right).
Both models have the same kinetic energy of the ejecta
($E = 1 \times 10^{51}$ ergs),
but the deposited energies are different
because of the difference in the gravitational potential energies
between the two models.
$Y_{e}^{\rm deep}$ is set to be 0.4985.
Note that a more massive supernova has more massive
complete and incomplete Si burning regions.
As one can see in Figure \ref{fig:pmdependence},
the mass ratio between the complete and incomplete Si-burning layers
depends on the progenitor mass and the mass cut.

\section{ABUNDANCE RATIOS IN METAL POOR STARS}\label{sec:abundhalo}

More massive stars evolve faster.
Thus, we expect the ejecta of the most massive stars
to dominate the earliest phase of Galactic evolution, i.e. the period
corresponding to the lowest [Fe/H].
If the mass cut $M_{\rm cut}$ between the ejecta and the neutron star
tends to be smaller for the larger mass progenitor
which provides earlier chemical input into the galactic evolution,
one could expect to reproduce the observed 
trend in [Cr/Fe], [Mn/Fe] and [Co/Fe].
Whether this trend agrees 
quantitatively with the observations has to be tested. 

The above trend between the progenitor mass and
the mass cut cannot be monotonic
because of two effects which compete with each other.
One is the amount of neutrino absorbing matter,
and the other is the depth of the gravitational potential.
For intermediate massive stars,
heavier stars eject relatively larger amounts of $^{56}$Ni
because of larger neutrino absorbing regions,
while in more massive stars, such as SN1997D,
the deeper gravitational potential wins and $^{56}$Ni is
scarcely ejected due to the fall back (Burrows 1998).
Note that recent observations, relating
the total $^{56}$Ni mass to progenitor mass,
seem to support this idea (see Figure \ref{fig:mvsni}).

We expect that the trend of increasing $^{56}$Ni yield
with increasing progenitor mass holds
over a large range of progenitor masses, while the reverse
trend occurs more abruptly.
Therefore, we focus here on the increasing trend
(decreasing mass cut with increasing progenitor mass)
up to a maximum progenitor mass,
above which we assume that SNe II do not contribute to galactic
chemical evolution.

\begin{figure}
\epsscale{.6}
\plotone{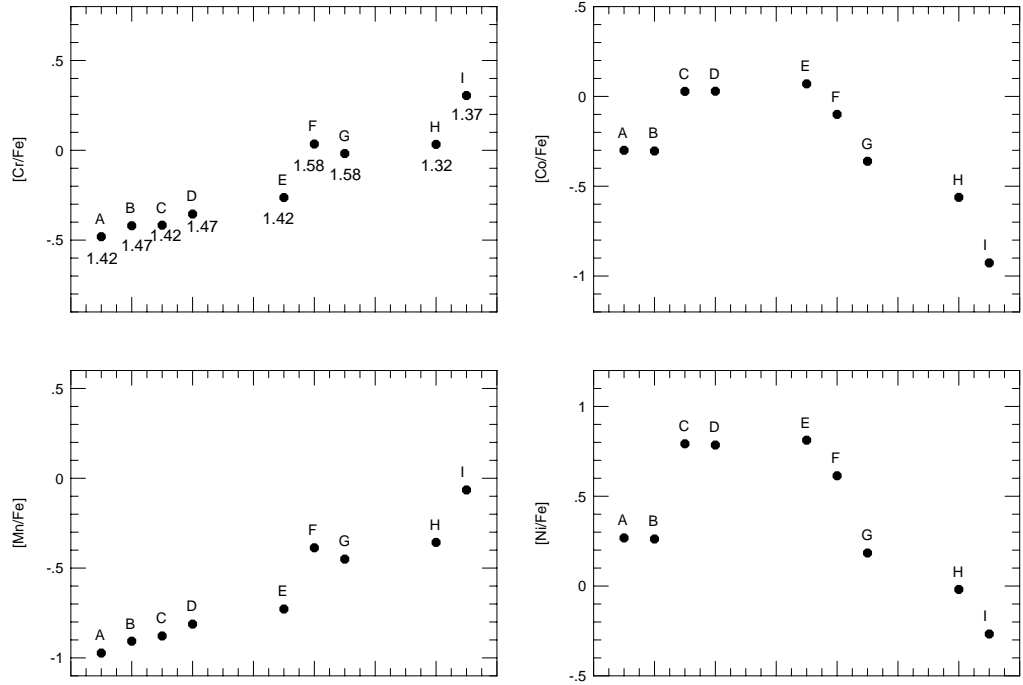}
\caption{
Abundance ratios of various SNe II.
Mass cuts of the models are also shown below abundance ratios.
Models {\bf A} and {\bf C} ({\bf B} and {\bf D}, {\bf F} and {\bf G})
have different $Y_{e}^{\rm deep}$ values (See Table \ref{tab:ratio4}).
\label{fig:ratio4}}
\end{figure}

\begin{figure}
\epsscale{.6}
\plotone{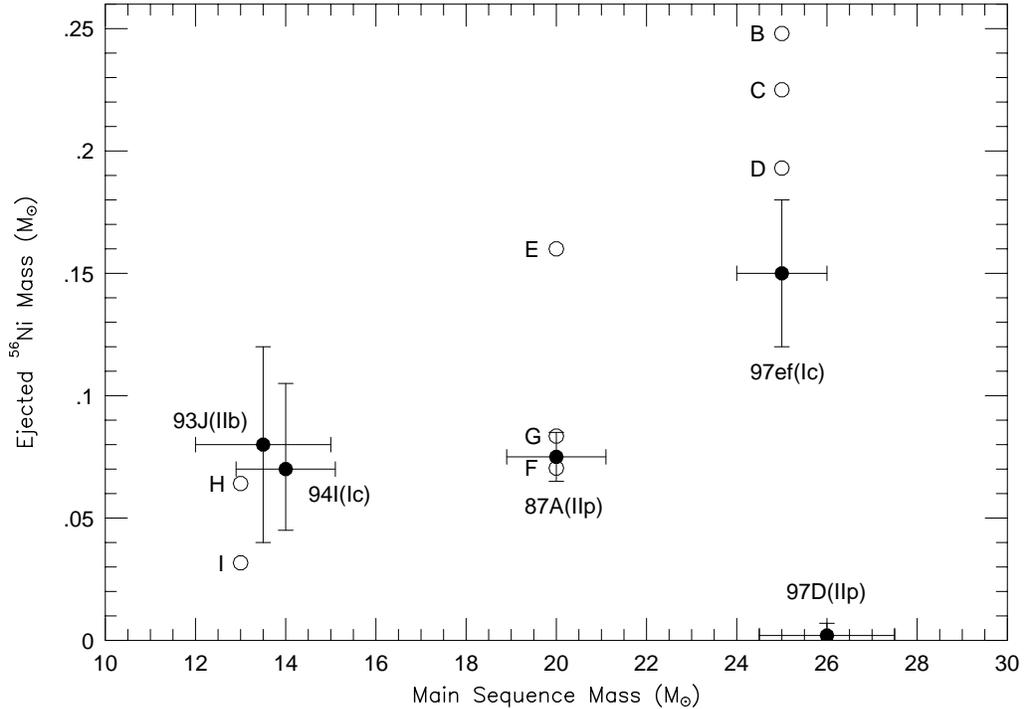}
\caption{
Ejected $^{56}$Ni mass as a function of main-sequence mass,
as estimated for SN II 1987A (Shigeyama \& Nomoto 1990);
SN IIb 1993J (Shigeyama et al. 1994);
SN Ic 1994I (Iwamoto et al. 1994);
SN II 1997D (Turatto et al. 1998);
SN Ic 1997ef (Iwamoto et al. 1998).
The $^{56}$Ni mass of the models in Figure \ref{fig:ratio4}
are shown as open circles.
\label{fig:mvsni}}
\end{figure}

Figure \ref{fig:ratio4} shows the abundance ratios of
various SN II models given in Table \ref{tab:ratio4}, where
their parameters (mass cut, etc) are summarized.
The abscissa of Figure \ref{fig:ratio4} indicates only
a model sequence (A-I).
For these models,
we take into account
the observational trend between the ejected $^{56}$Ni mass and stellar mass
by assuming that
the $^{56}$Ni yield of the 25$M_\odot$ star
is larger than that of less massive stars
(see Figure \ref{fig:mvsni}).
Correspondingly, Figure \ref{fig:ratio4} shows that deeper mass cuts yield
smaller [Cr/Fe] and [Mn/Fe], while they lead to larger [Co/Fe].
This means that if supernovae at smaller [Fe/H] have deeper mass cuts,
the observed trends can be explained.

We consider two models to relate [Fe/H] and the progenitor masses.
One is the `well-mixed' model, and the other is the 'unmixed' model.
In the 'well-mixed' model, we assume that the mixing
in the galactic scale is so efficient
that the chemical uniformity is achieved in the early Galaxy.
In the 'unmixed' model, on the contrary,
we assume that the mixing is not effective and
the composition of a metal-poor star is
the same as produced by a single SN II.

\placefigure{fig:ratio4}
\placetable{tab:ratio4}

\begin{figure}
\epsscale{.6}
\plotone{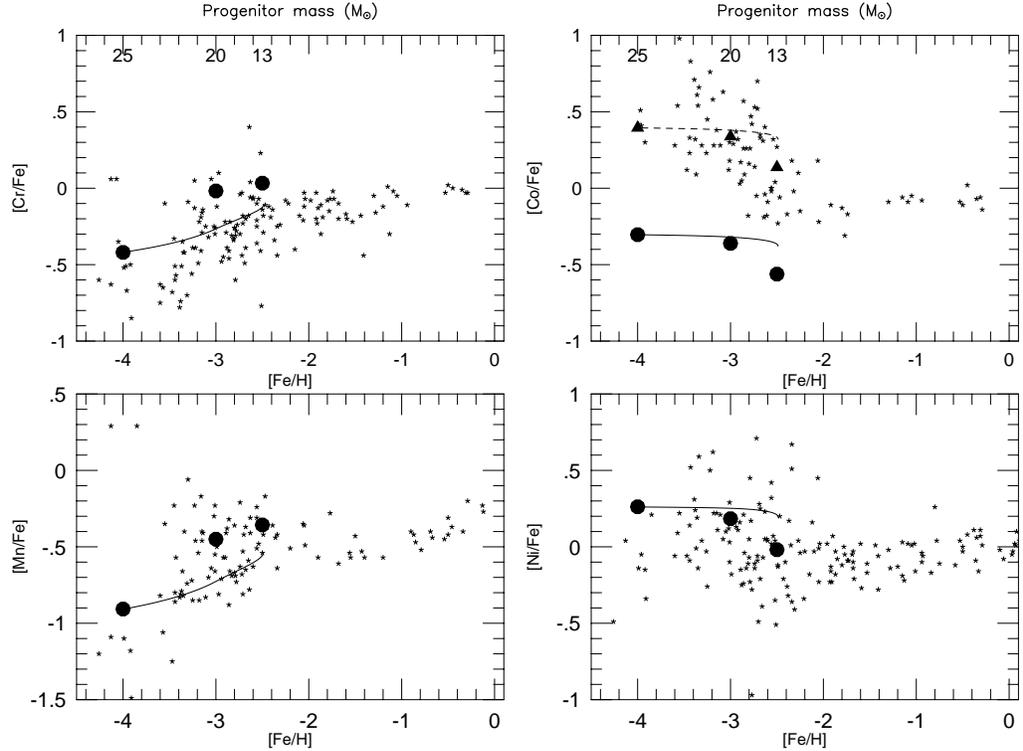}
\caption{Variations in the iron-peak abundance ratios.
The observational trend between ejected $^{56}$Ni mass and stellar mass
is taken into account.
Solid lines and dashed line show `well-mixed' models.
`Unmixed' models are also shown
in filled circles and filled triangles.
The model set [{\bf B}, {\bf G}, {\bf H}]
in Figure \ref{fig:ratio4} and Table \ref{tab:ratio4}
are used for obtaining these results.
The dashed line and the filled triangles indicate the models for which
the Co yield is multiplied by five
for easier comparison with observed data.
\label{fig:appli1}}
\end{figure}

For these models,
[Fe/H] is related to the stellar masses as follows.
For the 'well-mixed' model,
[Fe/H] at galactic time $t_G$,
as well as other abundances (e.g. Cu, Mn, and Co),
can be obtained by integrating the yields
from all SNe exploded for $t<t_G$.
[Fe/H] decreases monotonically towards the past
and the age at the lower [Fe/H] corresponds to
the shorter lifetime of more massive stars.
In this way, we construct the Galactic halo model
where the ages
at [Fe/H] $ = -4$, $ -3$ and $ -2.5$ roughly coincide with
the time when stars of 25$M_\odot$,
20$M_\odot$ and 13$M_\odot$ exploded, respectively.
Note that the mass vs. [Fe/H] relation is based on several assumptions
(star formation rate, initial mass function, etc.),
thus being subject to some uncertainties.
The solid curves in Figure \ref{fig:appli1} show the changes
in the iron-peak abundance ratios
as a function of [Fe/H].
To calculate these ratios, a set of
models [{\bf B}, {\bf G}, {\bf H}] is used
in Figure \ref{fig:ratio4},
and the Salpeter IMF of $\phi(M) \propto M^{-2.35}$ is adopted.
The observed trends of [Cr/Fe] and [Mn/Fe]
are relatively well reproduced
in Figure \ref{fig:appli1}.

In the 'unmixed' model,
the next generation stars were born in the ejecta of a SN II
which is uncontaminated by previous supernovae.
Then these stars have
the same heavy element abundances as a single SN II.
In this case, 
[Fe/H] depends on how much interstellar material is mixed
with the ejecta before forming the next generation stars.
Thus it is not trivial to see
why there is a relation between [Fe/H] and the progenitor mass $M$. 
In \S \ref{subsec:cbySNII}
we will discuss that the relation can be determined by
$M$-dependence of
the explosion energy and the Str\"omgren radius.
For now, we assign $M$ = 25, 20, and 13 $M_\odot$
to [Fe/H] = $-4, -3,$ and $-2.5$, respectively,
though the exact correspondence depends on
the $M$-dependent explosion energy and the Str\"omgren radius.
The filled circles in Figure \ref{fig:appli1}
show the 'unmixed' case
for the model set [{\bf B}, {\bf G}, {\bf H}].
The trends in [Cr/Fe] and [Mn/Fe] are well reproduced
with larger contrasts than in the 'well-mixed' model.

\begin{figure}
\epsscale{.6}
\plotone{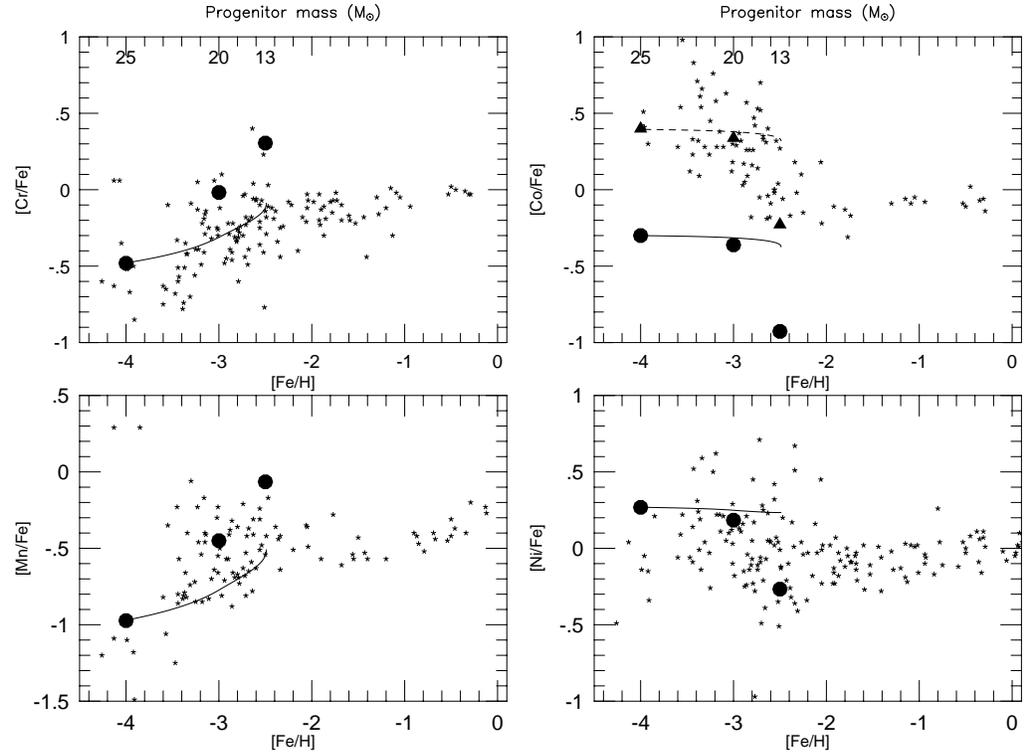}
\caption{
Same as Figure \ref {fig:appli1},
but with models for making stronger contrast in abundance ratios.
The model set [{\bf A}, {\bf G}, {\bf I}]
in Figure \ref{fig:ratio4} and Table \ref{tab:ratio4}
are used for the solid lines and the filled circles.
The dashed line and the filled triangles indicate the models for which
the Co yield is multiplied by five
for easier comparison with observed data.
\label{fig:appli2}}
\end{figure}

For other set of models,
the contrast in the abundance ratios can be
larger than the above case.
The solid lines and the filled circles in Figure \ref{fig:appli2}
use the set [{\bf A}, {\bf G}, {\bf I}],
and the contrasts in [Cr/Fe] and [Mn/Fe] are larger
than the set [{\bf B}, {\bf G}, {\bf H}]
for both the well-mixed and unmixed cases.

Though it is hard to argue which model,
between 'well-mixed' and 'unmixed',
is in better agreement
with the observations,
the slope of the abundance ratios for
the 'well-mixed' model seems to be slightly too gentle,
while the 'unmixed' model can produce a steeper slope.
Especially, the 'unmixed' model
better reproduces the inclination of the [Co/Fe] vs. [Fe/H] curve.

\begin{figure}
\epsscale{.6}
\plotone{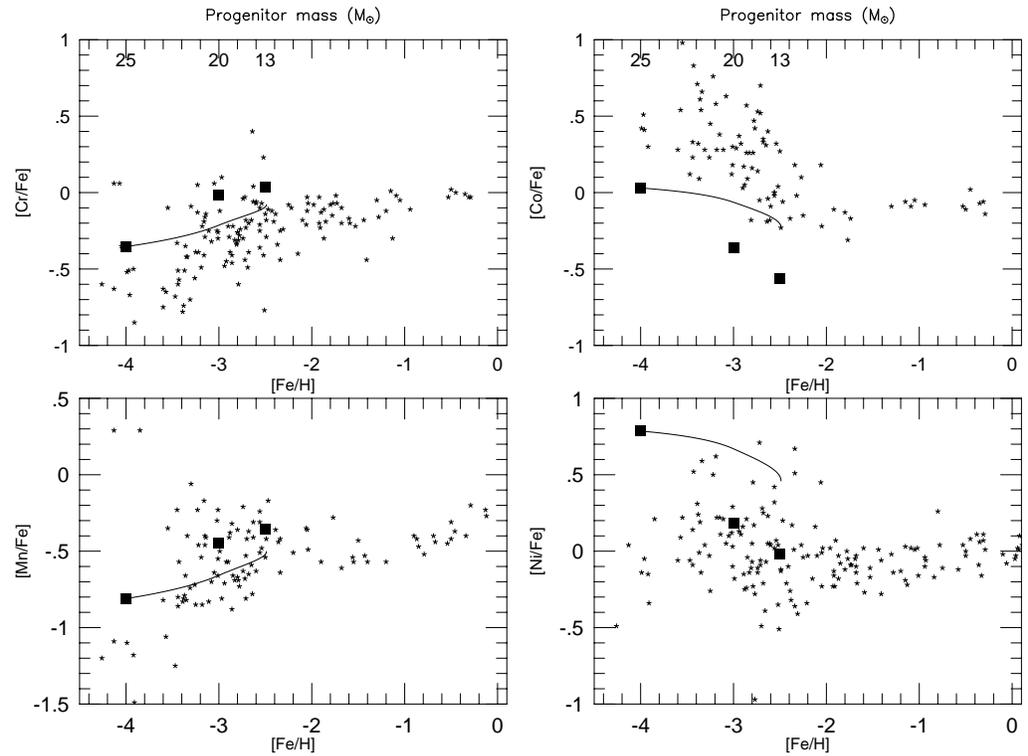}
\caption{Variations in the iron-peak abundance ratios.
The solid lines stand for `well-mixed' models
and the filled squares for `unmixed' models.
The model set [{\bf D}, {\bf G}, {\bf H}]
in Figure \ref{fig:ratio4} and Table \ref{tab:ratio4}
are used.
The model '{\bf D}' has smaller $Y_{e}^{\rm deep}$ than `{\bf B}'
\label{fig:appli3}}
\end{figure}

\begin{figure}
\epsscale{.6}
\plotone{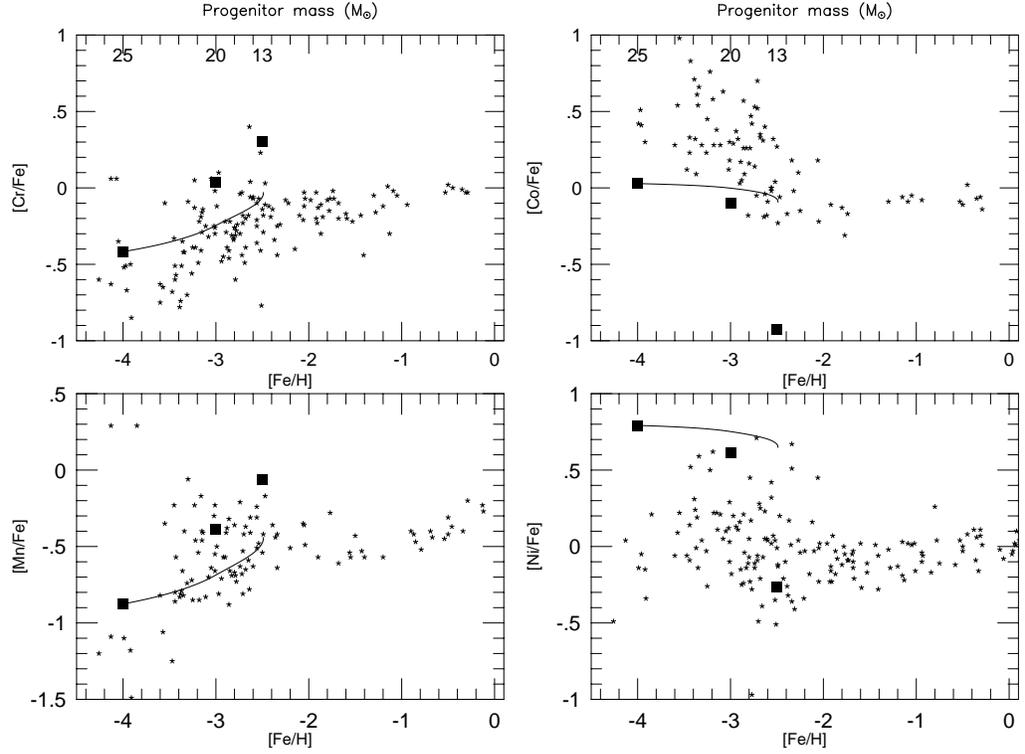}
\caption{
Same as Figure \ref {fig:appli2},
but with models for making stronger contrast in abundance ratios.
The models [{\bf C}, {\bf F}, {\bf I}]
in Figure \ref{fig:ratio4} and Table \ref{tab:ratio4}
are used.
\label{fig:appli4}}
\end{figure}

In Figures \ref{fig:appli1} and \ref{fig:appli2},
the calculated [Co/Fe] ratios are smaller
than the observed ratios by a factor of 3-5.
In order to compare the slope of the curve with
observations more easily,
the dashed lines in Figures \ref{fig:appli1} and \ref{fig:appli2}
show the models in which the amount of Co
is five times larger than
in the models
[{\bf B}, {\bf G}, {\bf H}] and
[{\bf A}, {\bf G}, {\bf I}].
The filled triangles in Figures \ref{fig:appli1} and \ref{fig:appli2}
indicate the five times enhanced Co for the 'unmixed' model.
As discussed in \S \ref{subsec:Yedep},
[Co/Fe] is larger for smaller $Y_{e}$.
Among the models in Figure \ref{fig:ratio4} and Table \ref{tab:ratio4},
'{\bf C}', '{\bf D}', and  '{\bf F}' have the same mass cut as
'{\bf A}', '{\bf B}', and  '{\bf G}', respectively,
but have smaller $Y_{e}$ and thus larger Co.
The solid lines for the 'well-mixed' model
and the filled squares for the `unmixed' model
in Figures \ref{fig:appli3} and \ref{fig:appli4}
show [{\bf D}, {\bf G}, {\bf H}]
and [{\bf C}, {\bf F}, {\bf I}], respectively.
Although [Co/Fe] is larger,
stable Ni ($^{58}$Ni, $^{60}$Ni, and $^{62}$Ni) is overproduced
and [Ni/Fe] is too large to be acceptable.
Therefore, the deficiency of $^{59}$Co (i.e., the deficiency of $^{59}$Cu)
is still an open question.

\placefigure{fig:appli1}
\placefigure{fig:appli2}
\placefigure{fig:appli3}
\placefigure{fig:appli4}

\section{DISCUSSION}\label{sec:discussions}
\subsection{Contamination by Type II supernovae in the very early Galaxy}\label{subsec:cbySNII}

In \S\ref{sec:abundhalo},
we have shown that the trends of [Cr/Fe] and [Mn/Fe] can be
well explained with both
the 'well-mixed' model and the 'unmixed' model.
However, the 'well-mixed' model 
tends to predict too small a metallicity dependence
in the abundances ratios.
Indeed, Audouze \& Silk (1995) and Ryan et al. (1996)
argue that only one, or at most
two to three, supernovae could have contaminated a particular cloud
in the [Fe/H] $ < -2.5$ region.
For the `unmixed' model,
the observed variations of [Cr/Fe], [Mn/Fe], and [Co/Fe] can be
explained with the systematic variations of these ratios as a function of
the progenitor mass $M$ (\S \ref{sec:abundhalo}).
The question is then
why there should be a tight correlation between [Fe/H] and $M$.

The [Fe/H] of a star is determined by the amount of iron
ejected from the relevant supernova and
the mass of hydrogen in the mixing region.
Our calculations show that in order to explain
the large variations of the observed abundance ratios
(e.g. $-0.5$ dex in [Cr/Fe]),
the iron mass from SN II
varies within a relatively narrow range of 
0.05$M_\odot$ to 0.25$M_\odot$.
Thus in the very low metallicity regions,
there must be 
an order of magnitude variation in the hydrogen mass
to produce the variation of
[Fe/H] in the range of $-4$ to $-2.5$.

Ryan et al. (1996) obtained
an analytic expression for
the mass of the ISM (interstellar matter), $m_{\rm ISM}$,
mixed with the ejecta using
equations (4.4b) and (3.33a) in Cioffi et al. (1988).
Ryan et al. (1996) suggested that
$m_{\rm ISM}$ depends only weakly on the environmental details,
but strongly on the explosion energy $E$ of the supernova
as $m_{\rm ISM} \propto E^{0.95}$. 
In this connection, we should note
the recent discovery of a supernova with such a large explosion energy
as $E \sim 3 \times 10^{52}$ erg
from a massive progenitor
(SN1998bw; Iwamoto et al. 1998).
If more massive supernovae tend to produce larger explosion energy
and this $M$-dependence is large enough,
the larger $M$ - smaller [Fe/H] relation can be obtained.

Here we propose another possibility.
The above discussion is derived under the assumption
that the ISM is uniform.
However, there is a distinctive non-uniformity characterized
by the ``Str\"omgren sphere'',
within which matter is ionized.
The progenitors of SNe II are so hot and luminous
during their main-sequence phase
that their Str\"omgren spheres are as large as $10 \sim 100$pc.
The Str\"omgren sphere is
likely to determine the amount of hydrogen into which
the ejecta of a supernova are mixed.
The shock advances easily within the Str\"omgren sphere,
but is strongly decelerated outside the sphere
because it has to ionize the matter there;
this means the effective adiabatic index $\gamma$ approaches $\gamma \sim 1$
outside the sphere.
The radius of the Str\"omgren sphere
is sensitive to the effective temperature of the star.
The change in the effective temperature by 25\%
leads to a doubling of the Str\"omgren radius (Osterbrock, 1989),
which results in a factor of $2^3 = 8$ change in the swept up hydrogen mass.
Therefore, it is possible to explain the observed range in [Fe/H].
Also the larger $M$ - smaller [Fe/H] relation
can be explained, because
a more massive, hotter star has a larger Str\"omgren radius.

This possibility can be corroborated by time scale estimates.
The recombination time scale, i.e.,
the Str\"omgren sphere's life time,
is $10^5$ - $10^6$ years.
The stellar evolution time after the main-sequence phase, plus
shock propagation time, is $10^5$ - $10^6$ years.
Therefore the Str\"omgren sphere produced during  the main-sequence phase
survives until the SN II explosion.
Then the shock radius coincides the Str\"omgren radius,
which depends on both the progenitor mass and
the number density of hydrogen
in the ambient ISM per cubic centimeter, $n_{0}$.
We estimate that the hydrogen mass swept up by the shock is
$4 \times 10^4 \cdot (n_{0}/1.0)^{-1} M_\odot $ for $M =$ 25$M_\odot$
and
$4 \times 10^3 \cdot (n_{0}/1.0)^{-1} M_\odot $ for $M =$ 15$M_\odot$,
using low metal progenitor models (Schaller et al. 1992).
These estimates show that
the hydrogen masses differ by a factor of 10.
However, we should note that,
for very small $n_{0}$ ($n_{0} \sim 0.1$),
the shock cannot reach the Str\"omgren radius,
so that the hydrogen mass does not depend on the Str\"omgren radius.
On the other hand, for larger $n_{0}$ ($n_{0} \sim 1$),
the Str\"omgren sphere vanishes before being approached by the shock.
In this case, however,
the difference in the swept up mass between larger and
smaller mass stars is larger,
because the longer life time of the latter makes it easier
for the Str\"omgren sphere to vanish.
This possibility is worth investigating.

\subsection{Mass of $^{56}$Ni
in Type II supernova ejecta}\label{subsec:mcinSNII}
Our results in \S \ref{sec:abundhalo} suggest that
a supernova from more massive progenitor has a deeper mass cut,
thus ejecting a larger amount of $^{56}$Ni mass.
The mass of $^{56}$Ni should be determined by the competition
between the amount of neutrino absorbing matter and 
the depth of the gravitational potential.
Accordingly, the intermediate massive stars eject
a relatively large amount of $^{56}$Ni
because of a large neutrino absorbing region,
while in a more massive star the deeper gravitational potential
wins and $^{56}$Ni is scarcely ejected due to fallback.

This is consistent with the recent $^{56}$Ni mass estimates
from the light curve modeling for SNe II
and Type Ib/Ic supernovae (SNe Ib/Ic)
which are the explosions of bare cores of massive stars
(Figure \ref{fig:mvsni}).
From SN IIb 1993J (Nomoto et al. 1993; Iwamoto et al. 1997),
SN Ic 1994I (Nomoto et al. 1994a; Iwamoto et al. 1994)
and SN1987A (e.g. Nomoto et al. 1994b for a recent review),
the 13 - 20 $M_\odot$ stars eject $\sim 0.07 M_\odot$ $^{56}$Ni.
For $M > 20 M_\odot$, there has been little information,
but recent SNe II/Ic suggest the strong mass-dependence of Fe yield.
The light curve of SN Ic 1997ef indicate the production of
$\sim 0.15 M_\odot$ of $^{56}$Ni (Iwamoto et al. 1998),
while that of SN II 1997D shows the synthesis of only
$\sim 0.002 M_\odot$ of $^{56}$Ni from the $25 - 30 M_\odot$ star
(Turatto et al. 1998).
SN II 1994W is also a small Fe producer (Sollerman et al. 1998).
Such an observed mass dependence of the $^{56}$Ni mass
is consistent with the theoretical expectation.

For  $M > 30 M_\odot$,
we expect that little $^{56}$Ni is ejected (WW95),
because of a large amount of fall back.
On the other hand,
the outer mantle including the oxygen-rich layer would be ejected
from such a massive supernova,
which is also required from
the Galactic chemical evolution model (Tsujimoto et al. 1997).
Possible contributions of recently discovered hypernovae
(Iwamoto et al. 1998) are worth investigating.

\placefigure{fig:mvsni}

\subsection{The Galactic chemical evolution}
In our new models,
stars around $\sim 25M_\odot$ have deeper mass cuts,
thus ejecting a larger amount of Fe
compared with the yields adopted by Tsujimoto et al. (1995).
It would be interesting to examine the consequences of
such a stellar-mass dependent Fe yield
for the Galactic chemical evolution model
and to compare with observations.
For [Fe/H] $\lsim -2.5$, however,
the error bars of the abundances in metal-poor stars
other than Cr, Mn, Co, Ni and Fe
are still too large to make meaningful comparisons (Ryan 1998).

For [Fe/H] $\gsim -2.5$, we can assume that
the ejecta from various SNe II are well mixed with ISM.
Thus the abundance ratios
between various elements and Fe are determined by
the total amount of Fe ejected by previous SNe II.
With the new choice of mass cuts,
the total amount of Fe is almost the same as
that in the previous model by Tsujimoto et al. (1995),
because Tsujimoto et al. (1995) assumed that
13-15 $M_\odot$ stars produce a larger amount of Fe
than stars more massive than 20$M_\odot$,
while in our model 13 - 20$M_\odot$ stars
produce smaller amounts of Fe than more massive stars.
These differences almost cancel out,
leading to similar total Fe yield from SNe II.

\begin{figure}
\epsscale{.6}
\plotone{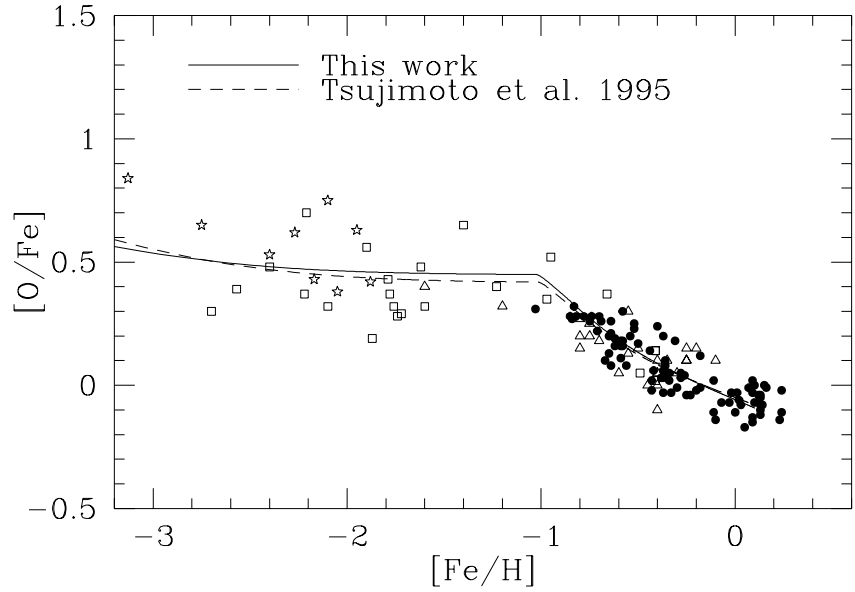}
\caption{
The evolution of [O/Fe] against [Fe/H]
for the different Fe yields from SNe II,
i.e., our model
with the set [{\bf B}, {\bf G}, {\bf H}]
(solid line)
and Tsujimoto et al. (1995; dashed line).
The observational data are taken from
Edvardsson et al. (1993) (filled circles);
Barbuy \& Erdelyi-Mendes (1989) (open triangles);
Nissen et al. (1994) (stars);
and Gratton (1991) (open squares).
\label{fig:oovfe}}
\end{figure}

Because the total Fe yield from SNe II is
similar to Tsujimoto et al. (1995),
the evolution of the abundance ratios such as [Cr/Fe] and [O/Fe] at
[Fe/H] $\gsim -2.5$ in our model is similar to
those in Tsujimoto et al. (1995).
Figure \ref{fig:oovfe} shows the evolution of [O/Fe] against [Fe/H]
for the different Fe yields from SNe II,
i.e., our model (solid line)
and Tsujimoto et al. (1995; dashed line).
In our model,
we use the set [{\bf B}, {\bf G}, {\bf H}]
and assume no Fe from stars more massive than 26$M_\odot$.
The upper mass and the lower mass of IMF
are set to be 40$M_\odot$ and 0.1$M_\odot$, respectively,
and the parameters of SNe Ia are chosen
to reproduce the abundance distribution
in the solar neighborhood (Kobayashi et al. 1998).
These parameters are only slightly different from
Tsujimoto et al. (1995).
Our model and Tsujimoto et al. (1995)
reproduce the observations almost equally well.
The evolution of the [Mn/Fe] ratio
between [Fe/H] = $-2.5$ and $-1.0$
depends more on
the SNe Ia rate rather than the SNe II rate (Kobayashi 1998).
At [Fe/H] $< -1$, [Mn/Fe] $\sim -0.5$ due to SNe II.
With increasing [Fe/H],
[Mn/Fe] approaches the solar value
because of the increasing contribution
of SNe Ia, which have [Mn/Fe] $ > 0$ according to
the SN Ia model W7 (Nomoto et al. 1984).
Detailed Galactic halo chemical evolution models
will be given in a separate paper.

\placefigure{fig:oovfe}

The deficiency of $^{59}$Co (i.e., $^{59}$Cu) remains a problem.
[Co/Fe] does not reach zero at [Fe/H] = 0
in the above Galactic chemical evolution model.
The most important parameter which affects the abundance of 
$^{59}$Cu is $Y_{e}^{\rm deep}$ (see \S\ref{subsec:Yedep}).
Lower $Y_{e}^{\rm deep}$ produces larger [Ni/Fe]
and [Co/Fe], which results in
a large deviation of [Ni/Fe] from the observations.
Therefore it seems unlikely that $Y_{e}$ solves this problem.
Could uncertainties in nuclear cross sections solve this problem?
This may not be easy because NSE is almost achieved
in the region that produces $^{59}$Cu.
Thus, cross sections would not have a large effect.
There are other elements which decay into Co,
but the amount of these elements is very small.
The deficiency of Co has to be investigated further.

In this work, we use progenitor models with solar metallicity.
Metallicity dependence may not affect our main conclusions
that the mass cut can explain the interesting trends
in the abundance ratios,
because the
production sites of the relevant elements and their
relative zone-thickness may not be so sensitive to metallicity.
However, some quantitative features,
e.g., the absolute value of each yield, may be
sensitive to metallicity.
Two of us (H. U. \& K. N.) are currently investigating
such metallicity effects.
There are other potentially important effects, such as
instabilities, dynamic convection, and rotation.
These effects also should be investigated in the future.

\bigskip

This work would not have been possible without the support of
Institute for Theoretical Physics, University of California,
Santa Barbara, USA,
itself supported under NSF grant no. PHY74-07194.
We would like to thank
Chiaki Kobayashi for preparing Figure \ref{fig:oovfe},
Raph Hix for providing us his nuclear
reaction network code,
Andrew McWilliam and Shigeru Kubono
for helpful suggestions and comments,
Sean Ryan for providing us with the data in his paper,
Masaaki Hashimoto for providing us with progenitor models,
and Chisato Ikuta for helping to gather observational data.
The authors acknowledge helpful discussions with
Toshikazu Shigeyama, Koichi Iwamoto and Chiaki Kobayashi
on several points in this paper.
Finally, we would like to thank the referee, Frank Timmes,
for useful comments to improve the paper.
This work has been supported by
the Swiss Nationalfonds grant (20-47252.96),
by the US NSF under grant AST 96-17494,
by the Grant-in-Aid for Scientific Research
(05242102, 06233101, 80203),
and by COE research (07CE2002)
of the Japanese Ministry of Education, Science, and Culture.

\clearpage

\clearpage

\begin{deluxetable}{crrrrrrc}
\footnotesize
\tablecaption{Dependence on mass cut.
\label{masscutdt}}
\tablewidth{0pt}
\tablehead{
\colhead{{\bf Model}}&\colhead{}&\colhead{}&\colhead{}
&\colhead{}&\colhead{}&\colhead{}&\colhead{}
}
\startdata
$M${\footnotesize ($M_\odot$)}&
25 & 25 & 25 & 20 & 20 & 20 &  \nl
$M_{{\rm core}}${\footnotesize ($M_\odot$)}&
8 & 8 & 8 & 6 & 6 & 6 &  \nl
E$_{{\rm exp}}{\footnotesize (\times 10^{51} {\rm ergs})}$
&$1.0$ & $1.0$
&$1.0$ & $1.0$
&$1.0$ & $1.0$
& \nl
${\bf M_{{\rm cut}}}${\footnotesize ($M_\odot$)}&
{\bf 1.42} & {\bf 1.54} & {\bf 1.62} &
{\bf 1.42} & {\bf 1.55} & {\bf 1.62} &  \nl
$Y_{e}^{\rm deep}$ & 0.4950 & 0.4950 & 0.4950 & 0.4940 & 0.4940 & 0.4940 & \nl
\hline
\hline
{\bf Yield {\footnotesize ($M_\odot$)}}&&&&&&& main elem. \nl
\hline
Fe& 2.44E-01 & 1.62E-01 & 1.14E-01 & 1.72E-01 & 9.53E-02 & 5.50E-02 & $^{56}$Ni\nl
Fe& 2.44E-01 & 1.62E-01 & 1.14E-01 & 1.72E-01 & 9.53E-02 & 5.50E-02 & $^{56}$Ni\nl
Cr& 1.30E-03 & 1.27E-03 & 1.24E-03 & 1.31E-03 & 1.20E-03 & 1.14E-03 & $^{52}$Fe\nl
Mn& 3.38E-04 & 3.37E-04 & 3.37E-04 & 3.35E-04 & 3.34E-04 & 3.33E-04 & $^{55}$Co\nl
Co& 6.87E-04 & 4.84E-04 & 3.01E-04 & 5.33E-04 & 2.33E-04 & 7.00E-05 & $^{59}$Cu\nl
Ni& 8.71E-02 & 5.48E-02 & 3.64E-02 & 6.42E-02 & 2.70E-02 & 7.16E-03 & $^{58}$Ni\nl
\hline
\hline
{\bf Ratio} &&&&&&& solar ratio\nl
\hline
Cr/Fe&
5.34E-03 & 7.87E-03 & 1.09E-02 & 7.62E-03 & 1.26E-02 & 2.08E-02 & 1.40E-02 \nl
[Cr/Fe]&
{\bf -0.417}&{\bf -0.249}&{\bf -0.106}&{\bf -0.263}&{\bf -0.044}&{\bf  0.173}&\nl
Mn/Fe&
1.38E-03 & 2.08E-03 & 2.96E-03 & 1.95E-03 & 3.50E-03 & 6.05E-03 & 1.04E-02 \nl
[Mn/Fe]& 
{\bf -0.878}&{\bf -0.700}&{\bf -0.548}&{\bf -0.728}&{\bf -0.475}&{\bf -0.273}&\nl
Co/Fe&
2.82E-03 & 2.77E-03 & 2.64E-03 & 3.10E-03 & 1.27E-03 & 2.45E-03 & 2.64E-03 \nl
[Co/Fe]& 
{\bf  0.028}&{\bf  0.022}&{\bf  0.001}&{\bf  0.070}&{\bf -0.032}&{\bf -0.316}&\nl
Ni/Fe&
3.57E-01 & 3.39E-01 & 3.20E-01 & 3.74E-01 & 2.83E-01 & 1.30E-01 & 5.76E-02 \nl
[Ni/Fe]& 
{\bf  0.792}&{\bf  0.770}&{\bf  0.745}&{\bf  0.812}&{\bf  0.692}&{\bf  0.355}&\nl
\enddata
\end{deluxetable}

\begin{deluxetable}{crrrrrrc}
\footnotesize
\tablecaption{Dependences on $Y_{e}$.
\label{yedt}}
\tablewidth{0pt}
\tablehead{
\colhead{{\bf Model}}&\colhead{}&\colhead{}&\colhead{}
&\colhead{}&\colhead{}&\colhead{}&\colhead{}
}
\startdata
$M${\footnotesize ($M_\odot$)}&
25 & 25 & 25 & 20 & 20 & 20 & \nl
$M_{{\rm core}}${\footnotesize ($M_\odot$)}&
8 & 8 & 8 & 6 & 6 & 6 & \nl
E$_{{\rm exp}}{\footnotesize (\times 10^{51} {\rm ergs})}$
&$1.0$ & $1.0$
&$1.0$ & $1.0$
&$1.0$ & $1.0$
&\nl
$M_{{\rm cut}}${\footnotesize ($M_\odot$)}&
1.42 & 1.42 &
1.42 & 1.42 &
1.42 & 1.42 & \nl
{\bf $Y_{e}^{\rm deep}$}& {\bf 0.4950} & {\bf 0.4965} & {\bf 0.4985} &
{\bf 0.4940} & {\bf 0.4965} & {\bf 0.4985} & \nl
\hline
\hline
{\bf Yield {\footnotesize ($M_\odot$)}}&&&&&&& main elem. \nl
\hline
Fe& 2.44E-01 & 2.71E-01 & 3.05E-01 & 1.72E-01 & 1.99E-01 & 2.21E-01 & $^{56}$Ni\nl
Cr& 1.03E-03 & 1.35E-03 & 1.41E-03 & 1.31E-03 & 1.37E-03 & 1.44E-03 & $^{52}$Fe\nl
Mn& 3.38E-04 & 3.38E-04 & 3.39E-04 & 3.35E-04 & 3.37E-04 & 3.38E-04 & $^{55}$Co\nl
Co& 6.87E-04 & 6.05E-04 & 4.03E-04 & 5.33E-04 & 4.47E-04 & 3.05E-04 & $^{59}$Cu\nl
Ni& 8.71E-02 & 6.31E-02 & 3.26E-02 & 6.42E-02 & 4.05E-02 & 2.29E-02 & $^{58}$Ni\nl
\hline
\hline
{\bf Ratio} &&&&&&& solar ratio\nl
\hline
Cr/Fe&
5.34E-03 & 4.97E-03 & 4.61E-03 & 7.62E-03 & 6.89E-03 & 6.51E-03 & 1.40E-02 \nl
[Cr/Fe]&
{\bf -0.417}&{\bf -0.448}&{\bf -0.481}&{\bf -0.263}&{\bf -0.306}&{\bf -0.331}&\nl
Mn/Fe&
1.38E-03 & 1.25E-03 & 1.11E-03 & 1.95E-03 & 1.69E-03 & 1.53E-03 & 1.04E-02 \nl
[Mn/Fe]& 
{\bf -0.878}&{\bf -0.922}&{\bf -0.973}&{\bf -0.728}&{\bf -0.790}&{\bf -0.833}&\nl
Co/Fe&
2.82E-03 & 2.34E-03 & 1.32E-03 & 3.10E-03 & 2.25E-03 & 1.38E-03 & 2.64E-03 \nl
[Co/Fe]& 
{\bf  0.028}&{\bf -0.072}&{\bf -0.300}&{\bf  0.070}&{\bf -0.070}&{\bf -0.281}&\nl
Ni/Fe&
3.57E-01 & 2.33E-01 & 1.07E-01 & 3.74E-01 & 2.03E-01 & 1.04E-01 & 5.76E-02 \nl
[Ni/Fe]& 
{\bf  0.792}&{\bf  0.607}&{\bf  0.268}&{\bf  0.812}&{\bf  0.548}&{\bf  0.256}&\nl
\enddata
\end{deluxetable}

\begin{deluxetable}{crrrrrrrrr}
\scriptsize
\tablecaption{Type II supernova Models in Figure \ref{fig:ratio4}.
\label{tab:ratio4}}
\tablewidth{0pt}
\tablehead{
\colhead{{\bf Model}}&
\colhead{A} & \colhead{B} & \colhead{C} &
\colhead{D} & \colhead{E} & \colhead{F} &
\colhead{G} & \colhead{H} & \colhead{I}
}
\startdata
$M${\footnotesize ($M_\odot$)}&
25 & 25 & 25 & 25 &
20 & 20 & 20 &
13 & 13 \nl
$M_{{\rm core}}${\footnotesize ($M_\odot$)}&
8  & 8   & 8   & 8   &
6  & 6   & 6   &
3.3& 3.3 \nl
E$_{{\rm exp}}{\footnotesize (\times 10^{51} {\rm ergs})}$
&$1.0$ & $1.0$&$1.0$&$1.0$&
 $1.0$ & $1.0$&$1.0$&
 $1.0$ & $1.0$\nl
$M_{{\rm cut}}${\footnotesize ($M_\odot$)}&
1.42 & 1.47 & 1.42 & 1.47 &
1.42 & 1.58 & 1.58 &
1.32 & 1.37 \nl
$Y_{e}^{\rm deep}$ &
0.4985 & 0.4985 & 0.4950 & 0.4950 &
0.4940 & 0.4940 & 0.4985 &
0.4990 & 0.4990 \nl
\hline
\hline
{\bf Yield {\footnotesize ($M_\odot$)}}&&&&&&&&& \nl
\hline
Fe& 3.05E-01 & 2.62E-01 & 2.44E-01 & 2.09E-01 &
    1.72E-01 & 7.77E-02 & 9.01E-02 &
    6.87E-02 & 3.50E-02 \nl
Cr& 1.41E-03 & 1.39E-03 & 1.30E-03 & 1.29E-03 &
    1.31E-03 & 1.18E-03 & 1.21E-03 &
    1.04E-03 & 9.85E-04 \nl
Mn& 3.40E-04 & 3.39E-04 & 3.38E-04 & 3.37E-04 &
    3.35E-04 & 3.33E-04 & 3.34E-04 &
    3.15E-04 & 3.14E-04 \nl
Co& 4.03E-04 & 3.43E-04 & 6.85E-04 & 5.89E-04 &
    5.33E-04 & 1.63E-04 & 1.03E-04 &
    4.97E-05 & 1.09E-05 \nl
Ni& 3.26E-02 & 2.75E-02 & 8.71E-02 & 7.36E-02 &
    6.42E-02 & 1.84E-02 & 7.92E-03 &
    3.79E-03 & 1.09E-03 \nl
$^{56}$Ni& 2.90E-01 & 2.48E-01 & 2.25E-01 & 1.93E-01 &
    1.60E-01 & 7.04E-02 & 8.35E-02 &
    6.41E-02 & 3.17E-02 \nl
\hline
\hline
{\bf Ratio} &&&&&&&&& \nl
\hline
[Cr/Fe]&
{\bf -0.481}&{\bf -0.420}&{\bf -0.418}&{\bf -0.355}&
{\bf -0.263}&{\bf  0.035}&{\bf -0.018}&
{\bf  0.033}&{\bf  0.305}\nl
[Mn/Fe]& 
{\bf -0.973}&{\bf -0.907}&{\bf -0.878}&{\bf -0.812}&
{\bf -0.728}&{\bf -0.387}&{\bf -0.450}&
{\bf -0.357}&{\bf -0.065}\nl
[Co/Fe]& 
{\bf -0.300}&{\bf -0.304}&{\bf  0.027}&{\bf  0.027}&
{\bf  0.070}&{\bf -0.100}&{\bf -0.361}&
{\bf -0.562}&{\bf -0.927}\nl
[Ni/Fe]& 
{\bf  0.268}&{\bf  0.262}&{\bf  0.792}&{\bf  0.785}&
{\bf  0.812}&{\bf  0.614}&{\bf  0.184}&
{\bf -0.019}&{\bf -0.267}\nl
\enddata
\end{deluxetable}

\end{document}